\documentclass[aps,twocolumn,nofootinbib,prd,aps,10pt]{revtex4}
\usepackage[dvips]{graphicx}
\usepackage[english]{babel}
\usepackage[T1]{fontenc}
\usepackage{mathrsfs}
\usepackage[tbtags]{amsmath}
\usepackage{amssymb}
\usepackage{amsxtra}
\usepackage{amsopn}
\usepackage{latexsym}
\usepackage[mathcal]{eucal}
\usepackage{mathtools}

\newcommand{\BE}{\begin{equation}}
\newcommand{\EE}{\end{equation}}
\newcommand{\BA}{\begin{align}}
\newcommand{\EA}{\end{align}}
\newcommand{\Tr}{\mathrm Tr}
\newcommand{\nn}{\nonumber}

\begin{document}

\title{Calculation of the non-perturbative strong coupling from first principles}

\author{Fabio Siringo}

\affiliation{Dipartimento di Fisica e Astronomia 
dell'Universit\`a di Catania,\\ 
INFN Sezione di Catania,
Via S.Sofia 64, I-95123 Catania, Italy}

\date{\today}

\begin{abstract}
The success of the screened massive expansion is investigated in the framework of a screened 
momentum-subtraction scheme for the running of the strong coupling in pure Yang-Mills theory.
By the exact Slavnov-Taylor and Nielsen identities, a very predictive and self-contained set of
stationary conditions are derived for the optimization of the fixed-coupling expansion, 
yielding explicit analytical one-loop expressions for the propagators, the coupling and the beta function,
from first principles. An excellent agreement is found with the lattice data. In the proposed
screened renormalization scheme, a monotonic running coupling emerges which saturates in the IR 
at the finite IR stable fixed point $g=9.40$ where the beta function crosses the zero. A simple
analytical expression is derived for the leading behavior of the beta in the IR.
\end{abstract}




\maketitle

\section{Introduction}

In the last decades, considerable progress has been made in the study of the non-perturbative low-energy regime of
QCD and pure Yang-Mills theory. Because of the breakdown of ordinary perturbation theory (PT), most of our knowledge
on confinement, hadron mass spectra and mass generation\cite{cornwall,bernard,dono,philip,aguilar04,papa15b,olive} 
arises from lattice calculations\cite{olive,cucch07,cucch08,cucch08b,cucch09,bogolubsky,olive09,dudal,binosi12,olive12,burgio15,duarte} 
or from the numerical solution of integral equations in the continuum\cite{aguilar8,aguilar10,aguilar14,papa15,fischer2009,
huber14,huber15g,huber15b,pawlowski08,pawlowski10,pawlowski10b,pawlowski13,
varqcd,genself,highord,watson10,watson12,rojas,reinhardt04,reinhardt05,reinhardt14}.
Only few analytical results have been reached, by phenomenological models,
based on modified quantization procedures or different Lagrangians\cite{GZ,dudal08,dudal08b,dudal11,tissier10,tissier11,serreau}.

In the last years, a purely analytical approach to the exact gauge-fixed Lagrangian of QCD has been developed 
by a screened massive expansion\cite{ptqcd,ptqcd2,scaling,analyt,xigauge,damp}.
Preliminary variational calculations\cite{varqcd,genself,highord} had shown that PT might actually work at low energy and give reasonable
results at the lowest orders of approximation if the expansion point is changed and the expansion is taken around a massive 
gluon propagator. Regardless of the accuracy of the zeroth order gluon propagator, provided that it has a mass scale, the higher-order terms
become small, suggesting that the failure of PT might be a consequence of the bad choice of expanding around the usual mass-less 
gluon propagator of the high energy theory. On the other hand, 
the even earlier remarkable discovery of Tissier and Wschebor\cite{tissier10,tissier11},
that PT is viable if a gluon mass is added by hand to the Lagrangian, suggested that the best expansion point might actually be
the simple free massive gluon which emerges by just adding a gluon mass term to the quadratic part of the Lagrangian and subtracting it again
from the interaction, thus leaving the total action unchanged. Expanding around the modified massive quadratic part of the action, leads to
the screened massive expansion which was developed in Refs.\cite{ptqcd,ptqcd2} and has the merit of providing explicit and very
accurate analytical expressions for the propagators of the gauge-fixed Faddeev-Popov Lagrangian, without adding any phenomenological parameter\cite{scaling}.
The method was extended to the full QCD in Ref.\cite{analyt}, to a generic covariant gauge in Ref.\cite{xigauge} 
and to finite temperature in Refs.\cite{damp,varT}, 
enforcing the idea that most of the non-perturbative effects can be embedded in the gluon-mass parameter\cite{tissierplb,tissiersu2,tissier15,tissier16}.

A great advantage of the screened expansion is that, when optimized, it provides explicit analytical expressions which can be continued
to the whole complex plane, gaining access to information that cannot be currently obtained 
by other theoretical tools\cite{dudalolive14,dudalolive19,kondo18,dispersion}.
At variance with the Curci-Ferrari model which was studied in Refs.\cite{tissier10,tissier11} as
a low energy effective model, in the screened expansion no mass term is added to 
the total action which is not modified at all and still has the usual exact Becchi-Rouet-Stora-Tyutin (BRST) symmetry.
It is the modified expansion point which breaks the symmetry softly at any finite order, so that the exact constraints inherited by the BRST symmetry,
like Slavnov-Taylor (ST) and Nielsen identities\cite{nielsen}, 
can be used as a self-contained criterion for optimizing the expansion and fixing the finite part of
the renormalization constants, which are usually scheme-dependent and undetermined in PT. Without any phenomenological parameter left, the optimized
expansion is highly predictive, as shown in Ref.\cite{xigauge} where the optimal gluon propagator was determined by the Nielsen identities.

In this paper, the optimization of the screened expansion for pure Yang-Mills theory is revised and extended
to the ghost propagator, which plays a key role for 
determining the running coupling in the Landau gauge\cite{taylor,alkofer}. 
A natural variant of the momentum-subtraction (MOM) scheme
is proposed by requiring that, at  the subtraction point,  the renormalized gluon propagator is equal to the screened massive free-propagator instead
of the massless propagator of standard PT. In that screened scheme, the expansion is optimized using Nielsen identities, ST identities and
enforcing the multiplicative renormalization of the theory. From first principles, without any phenomenological parameter, the gluon and ghost
propagators are found in excellent agreement with the lattice data. Moreover an {\it ab initio} lattice-independent measure of the 
strong coupling is provided in the continuum, avoiding the effects of a finite lattice spacing\cite{duarte}.

In the screened MOM scheme, the strong coupling is a monotonic function which saturates in the IR at a finite fixed point where the beta function
meets a zero. Thus, the proposed scheme could have more phenomenological relevance than the standard MOM scheme.
Furthermore, analytical expressions can be easily derived for the strong coupling and the beta as functions of the renormalization scale.
The non-perturbative beta function is fully characterized and a simple 
explicit expression is provided for the leading behavior at the finite IR fixed point.

The paper is organized as follows:
in Section II, the renormalization of Yang-Mills theory is briefly reviewed  in order 
to fix the notation and recall some general results; in Section III a screened MOM
scheme is introduced for the screened massive expansion; in Section IV the fixed-coupling
expansion is optimized by a set of stationary conditions which derive from the BRST symmetry
of the total action; in Section V the predictions of the optimized expansion are
compared with the available lattice data and the properties of the beta function are described in detail;
finally, a brief discussion of the main results is given in Section VI. Explicit analytical expressions and
the leading behavior of the propagators, strong coupling and beta function are collected in Appendix A.

\section{Renormalization of Yang-Mills theory}

In a linear covariant $\xi$-gauge, the exact gauge-fixed bare Lagrangian of pure Yang-Mills  SU(N) theory can be written as
\BE
{\cal L}={\cal L}_{YM}+{\cal L}_{fix}+{\cal L}_{FP}
\label{Ltot}
\EE
where 
\begin{align}
{\cal L}_{YM}&=-\frac{1}{2} \Tr\left(  \hat F_{\mu\nu}\hat F^{\mu\nu}\right),\nn\\
{\cal L}_{fix}&=-\frac{1}{\xi} \Tr\left[(\partial_\mu \hat A^\mu)(\partial_\nu \hat A^\nu)\right]
\label{LYMfix}
\end{align}
and ${\cal L}_{FP}$ is the ghost term
arising from the Faddeev-Popov (FP) determinant.
Here, the tensor operator is
\BE
\hat F_{\mu\nu}=\partial_\mu \hat A_\nu-\partial_\nu \hat A_\mu
-i g \left[\hat A_\mu, \hat A_\nu\right],
\label{F}
\EE
and the gauge field operators satisfy the $SU(N)$ algebra
\begin{align}
\hat A^\mu&=\sum_{a} \hat X_a A_a^\mu\nn\\
\left[ \hat X_a, \hat X_b\right]&= i f_{abc} \hat X_c,\quad
f_{abc} f_{dbc}= N\delta_{ad}.
\end{align}

The total action can be split as $S_{tot}=S_0+S_I$ where the quadratic part is
\begin{align}
S_0&=\frac{1}{2}\int A_{a\mu}(x)\delta_{ab} {\Delta_0^{-1}}^{\mu\nu}(x,y) A_{b\nu}(y) {\rm d}^4 x\,{\rm d}^4 y \nn \\
&+\int c^\star_a(x) \delta_{ab}{{\cal G}_0^{-1}}(x,y) c_b (y) {\rm d}^4 x\, {\rm d}^4 y
\label{S0}
\end{align}
while the interaction contains the three vertices
\BE
S_I=\int{\rm d}^4x \left[ {\cal L}_{gh} + {\cal L}_3 +   {\cal L}_4\right],
\label{SI}
\EE
\begin{align}
{\cal L}_{3g}&=-g  f_{abc} (\partial_\mu A_{a\nu}) A_b^\mu A_c^\nu\nn\\
{\cal L}_{4g}&=-\frac{1}{4}g^2 f_{abc} f_{ade} A_{b\mu} A_{c\nu} A_d^\mu A_e^\nu\nn\\
{\cal L}_{ccg}&=-g f_{abc} (\partial_\mu c^\star_a)c_b A_c^\mu.
\label{Lint}
\end{align}
In Eq.(\ref{S0}), $\Delta_0$ and ${\cal G}_0$ are the standard free-particle propagators for
gluons and ghosts and their Fourier transforms are
\begin{align}
{\Delta_0}^{\mu\nu} (p)&=\Delta_0(p)\left[t^{\mu\nu}(p)  
+\xi \ell^{\mu\nu}(p) \right]\nn\\
\Delta_0(p)&=\frac{1}{-p^2}, \qquad {{\cal G}_0} (p)=\frac{1}{p^2},
\label{D0}
\end{align}
having used the transverse and longitudinal projectors 
\BE
t_{\mu\nu} (p)=g_{\mu\nu}  - \frac{p_\mu p_\nu}{p^2};\quad
\ell_{\mu\nu} (p)=\frac{p_\mu p_\nu}{p^2}.
\label{tl}
\EE

In the present paper, we will take $\xi\to 0$ and use the Landau gauge which is a Renormalization Group (RG) fixed point and
is the most studied gauge on the lattice. Moreover, we will take advantage of Taylor's non-renormalization theorem\cite{taylor} which is only
valid in the Landau gauge. Therefore, we will not bother about the gauge parameter $\xi$ in the following discussion.

In the above equations, the fields and the coupling must be regarded as the bare $A_B$, $c_B$, $g_B$. 
The total Lagrangian in Eq.(\ref{Ltot}) has
a BRST symmetry which can be used for proving exact Slavnov-Taylor (ST) identities,  Nielsen identities,
and the multiplicative renormalizability of the  theory.
Then, we can follow the standard path of renormalization and introduce usual renormalized fields and coupling $A$, $c$, $g$
\begin{align}
A^\mu_B&=\sqrt{Z_A}\> A^\mu, \quad g_B=Z_g\> g\nn\\
c_B&=\sqrt{Z_c}\> c,\qquad c^\star_B=\sqrt{Z_c}\> c^\star
\end{align}
having dropped obvious color indices. In terms of the renormalized fields, the Lagrangian can be written in the same identical form as 
in Eq.(\ref{Ltot}), provided that a set of counterterms are added to the interactions of Eq.(\ref{Lint}). For instance, with an obvious
shorthand notation, the gluon quadratic part becomes
\BE
S_0^g=\frac{1}{2}\int A_{\mu}\> {\Delta_0^{-1}}^{\mu\nu} A_{\nu}+ \frac{1}{2}\delta Z_A\int A_{\mu}\> {\Delta_0^{-1}}^{\mu\nu} A_{\nu}
\label{counterA}
\EE
where $\delta Z_A=(Z_A-1)\sim {\cal O}(g^2)$, so that the second term on the right hand side can be regarded as a counterterm which contributes 
to the polarization at tree-level. On the same footing, the gluon-ghost vertex ${\cal L}_{ccg}$ can be written as
\begin{align}
{\cal L}_{ccg}&=-Z_1^c\> g f_{abc} (\partial_\mu c^\star_a)c_b A_c^\mu=\nn\\
&=- g f_{abc} (\partial_\mu c^\star_a)c_b A_c^\mu-\delta Z_1^c\> g f_{abc} (\partial_\mu c^\star_a)c_b A_c^\mu
\label{counter1}
\end{align}
where the vertex renormalization constant $Z_1^c$ is defined as
\BE
Z_1^c=Z_g\, Z_c\,\sqrt{Z_A}
\label{Z1c}
\EE
and $\delta Z_1^c=(Z_1^c-1)$ is the prefactor of the vertex counterterm in the last term of Eq.(\ref{counter1}).

An important consequence of ST identities is that all divergences can be eliminated by the same $Z_g$ for all vertices,
since the ghost-gluon, three-gluon and four-gluon vertex renormalization constants, $Z_1^c$, $Z_1^{3g}$ and $Z_1^{4g}$ respectively, 
must satisfy the exact equations
\BE
\frac{Z_1^c}{Z_c}=\frac{Z_1^{\>3g}}{Z_A}=\left(\frac{Z_1^{\>4g}}{Z_A}\right)^{\frac{1}{2}}.
\label{Zs}
\EE

In perturbation theory (PT), the diverging parts of the renormalization constants satisfy Eqs.(\ref{Zs}) at each order, but their finite
parts might not respect the same constraints. Actually, in PT the finite parts of the renormalization constants are not determined at all
and depend on the renormalization scheme. While they are simply ignored in the Minimal Subtraction scheme  ($\overline{MS}$), a more physical 
criterion is provided by the Momentum Subtraction scheme (MOM). The optimization of that choice leads to a better convergence of PT if the
finite parts can be used as variational parameters, yielding an optimized PT by variation of the renormalization scheme\cite{stevensonRS}.
Of course, the use of that method requires the existence of a function which is known to be minimal when the approximate result approaches
the exact one, or at least stationary, according to Stevenson's principle of minimal sensitivity\cite{sensitivity}.

As discussed in Ref.\cite{xigauge}, one important merit of the screened massive expansion arises from the apparent
drawback of having split the action in two parts that are not BRST invariant. The exact constraints arising from BRST invariance
cannot be satisfied exactly at any finite order of the expansion, but they must be satisfied by the exact result. Thus,
ST and Nielsen identities provide a criterion for optimizing the expansion and fixing the finite part of the renormalization constants.
In Ref.\cite{xigauge} the optimal gluon propagator was determined by requiring that the poles and the phases of the residues are gauge
parameter independent, as required by Nielsen identities. At one-loop, the method provides a very accurate analytical expression for
the gluon propagator, in very good agreement with the lattice data.

The method can be extended to the ghost propagator, which plays a key role for determining the running coupling because of
the non-renormalization of the ghost-gluon vertex in the Landau gauge. The extension must rely on ST identities and 
on the multiplicative renormalization of the theory, according to Eqs.(\ref{Zs}). 

\section{Screened Momentum-Subtraction}

A screened expansion for the exact renormalized Yang-Mills (YM) Lagrangian was developed in Refs.\cite{ptqcd,ptqcd2} and  extended later
to finite temperature in Refs.\cite{damp,varT} and to the full QCD in Ref.\cite{analyt}, where a set of chiral quarks were included. 
The extension to a generic covariant gauge\cite{xigauge} has already shown the predictive power of the method 
when the expansion is optimized by the constraints of BRST symmetry.

The screened expansion arises by changing the expansion point {\it after} having renormalized the fields and the coupling 
as discussed above. Thus, the exact renormalization constants still satisfy Eqs.(\ref{Zs}) because of the ST identities.

Following Refs.\cite{ptqcd2,xigauge} the new screened expansion can be recovered by just adding a transverse mass term to the quadratic
part of the action and subtracting it again from the interaction, leaving the total action
unchanged. After having renormalized the fields and the coupling, we add and subtract the action term
\BE
\delta S= \frac{1}{2}\int A_{a\mu}(x)\>\delta_{ab}\> \delta\Gamma^{\mu\nu}(x,y)\>
A_{b\nu}(y) {\rm d}^4\, x{\rm d}^4y
\label{dS1}
\EE
where the vertex function $\delta\Gamma$ is a shift of the inverse propagator
\BE
\delta \Gamma^{\mu\nu}(x,y)=
\left[{\Delta_m^{-1}}^{\mu\nu}(x,y)- {\Delta_0^{-1}}^{\mu\nu}(x,y)\right]
\label{dG}
\EE
and ${\Delta_m}^{\mu\nu}$ is a new massive free-particle propagator 
\begin{align}
{\Delta_m^{-1}}^{\mu\nu} (p)&=
(-p^2+m^2)\,t^{\mu\nu}(p)  
+\frac{-p^2}{\xi}\ell^{\mu\nu}(p).
\label{Deltam}
\end{align}
Adding that term is equivalent to substituting the new massive propagator ${\Delta_m}^{\mu\nu}$ for the 
old massless one ${\Delta_0}^{\mu\nu}$ in the quadratic part. Thus, the new expansion point is a massive
free-particle propagator for the gluon, which is much closer to the exact propagator in the IR.

In order to leave the total action unaffected by the change, the same term is added in the interaction,
providing a new interaction vertex $\delta\Gamma$, a two-point vertex which can be regarded as a new counterterm.
Dropping all color indices in the diagonal matrices and
inserting Eq.(\ref{D0}) and (\ref{Deltam}) in Eq.(\ref{dG}) the vertex is just the transverse mass shift
of the quadratic part
\BE
\delta \Gamma^{\mu\nu} (p)=m^2 t^{\mu\nu}(p) 
\label{dG2}
\EE
and must be added to the standard set of vertices arising from Eq.(\ref{Lint}). The new vertex does not contain
any renormalization constant and is part of the interaction even if it does not depend on the coupling. 
The expansion must be regarded as a $\delta$-expansion more than a loop expansion, since different powers of the coupling
coexist at each order in powers of the total interaction.
Then, the proper gluon polarization and ghost self energy can be evaluated, order by order, by PT.
In all Feynman graphs, the internal gluon lines are replaced by the massive free-particle propagator ${\Delta_m}^{\mu\nu}$ and 
new insertions must be considered of the (transverse) two-point vertex $\delta \Gamma^{\mu\nu}$. 
For further details of the screened expansion we refer to Refs.\cite{ptqcd2,xigauge} where explicit analytical expressions are reported up
to third order in the $\delta$-expansion and one-loop.

Since the total gauge-fixed FP Lagrangian is not modified and because of gauge invariance,
the longitudinal polarization is known exactly and is zero. 
We can write the exact polarization as
\BE
\Pi^{\mu\nu}(p)=\Pi(p)\, t^{\mu\nu}(p)
\label{pol}
\EE
so that, in the Landau gauge, the exact gluon propagator is transverse 
\BE
\Delta_{\mu\nu}(p)=\Delta (p)\,t_{\mu\nu}(p)
\EE
and defined by the single scalar function $\Delta(p)$.

From now on we switch to the Euclidean formalism and denote by $p^2$ the Euclidean
square which is {\it negative} on the positive real axis of Minkowski space.
The exact (dressed) gluon and ghost propagators are given by the functions
\begin{align}
{\Delta}^{-1} (p)&=p^2+m^2-\Pi(p)\nn\\
{\cal G}^{-1}(p)&=-p^2-\Sigma (p)
\label{dressprop}
\end{align}
where the proper gluon polarization $\Pi$ and ghost self-energy $\Sigma$ are the
sum of all one-particle-irreducible (1PI) graphs in the screened expansion, including all counterterms.

We observe that the mass parameter $m$ is arbitrary and a renormalization of $m$ does not make any sense, 
since it is an arbitrary quantity which is added and subtracted again in the action. It will be regarded
as a RG invariant.
On the other hand, BRST invariance protects the expansion from the appearing of any spurious diverging mass term, so that
no mass counterterm is required and no diverging renormalization constant would require a renormalization of $m$.
Actually, the added mass term breaks the BRST symmetry of the quadratic part $S_0$ and of the interaction $S_I$ when taken apart.
Therefore, many constraints arising from BRST are not satisfied exactly at any finite order of the screened expansion.
While the soft breaking has no effect on the UV behavior and on the diverging parts of the renormalization constants, some spurious
diverging mass terms do appear in the expansion at some stage. However, as discussed in Refs.\cite{ptqcd2,analyt,xigauge}, the insertions of
the new vertex $\delta \Gamma$, Eq.(\ref{dG2}), cancel the spurious divergences exactly, without the need of any mass renormalization,
as a consequence of the unbroken BRST symmetry of the whole action $S_0+S_I$.
That aspect makes the screened expansion very different from effective models like the Curci-Ferrari model where a bare mass term is 
present in the Lagrangian from the beginning.

The exact self energies can be written as
\begin{align}
\Pi(p)&=m^2-p^2\delta Z_A+\Pi_{loop}(p)\nn\\
\Sigma (p)&=p^2\delta Z_c+\Sigma_{loop}(p)
\label{selfs}
\end{align}
where the tree-level contribution $m^2$ comes from the new two-point vertex $\delta \Gamma$ in Eq.(\ref{dG2}), while the tree-level
terms $-p^2\delta Z_A$, $p^2\delta Z_c$ arise from the respective field-strength renormalization counterterm, which was explicitly shown 
in Eq.(\ref{counterA}) for the gluon. The proper functions $\Pi_{loop}$, $\Sigma_{loop}$ are given by the sum of all 1PI graphs containing
loops.

The diverging parts of $\delta Z_A$, $\delta Z_c$ cancel the UV divergences of $\Pi_{loop}$ and $\Sigma_{loop}$, respectively. 
Since these divergences do
not depend on masses, they are exactly the same as in the standard PT, so that in the $\overline{MS}$ scheme $Z_A$ and $Z_c$ have their standard
expressions, as manifest in the explicit one-loop calculation\cite{ptqcd2}. 
The finite parts of $\delta Z_A$, $\delta Z_c$ are arbitrary and depend on the scheme. On the other hand, the self energies contain an arbitrary
term $Cp^2$, where $C$ is a constant which depends on the regularization method. 

The MOM scheme is usually regarded as the most physical way to fix the 
finite parts for QCD in the Landau gauge. However, for the screened expansion, 
the  most appropriate scheme seems to be a screened version\cite{tissier11} of the
standard MOM, which we call screened-MOM (SMOM). As suggested by Eq.(\ref{dressprop}), the SMOM scheme is defined by
requiring that the dressed propagators are equal to the free-propagators of the screened expansion at the scale $p=\mu$:
\begin{align}
{\Delta}^{-1} (\mu)&=\mu^2+m^2\nn\\
{\cal G}^{-1}(\mu)&=-\mu^2,
\label{SMOM1}
\end{align}
at variance with the usual MOM scheme where the gluon propagator is set as $\Delta(\mu)=1/\mu^2$.
We observe that the validity of Taylor's non-renormalization theorem\cite{taylor} is not affected by the masses and still holds in the screened expansion
and in the Curci-Ferrari model\cite{wschebor,tissiervertex}. Thus, we can set $Z_1^c=1$ and require that the finite part of $Z_g$ satisfies the exact Eq.(\ref{Z1c})
which completes the definition of the scheme
\begin{align}
Z_1^c&=1 \qquad (\xi=0)\nn\\
Z_g^{-1}&=Z_c\sqrt{Z_A}.
\label{SMOM2}
\end{align}
In Ref.\cite{tissier11}, Eq.(\ref{SMOM1}) was used in a MOM scheme for the Curci-Ferrari model, but Eq.(\ref{SMOM2}) was different
because the mass was a running coupling while here $m$ is a RG invariant.

The SMOM scheme is non-perturbative and Eqs.(\ref{SMOM1}),(\ref{SMOM2}) are exact
equations governing the multiplicative renormalization of the exact dressed propagators in
Eq.(\ref{dressprop}). The approximate propagators that arise from the screened expansion 
might not be fully compatible with Eq.(\ref{SMOM2}) which is expected to be satisfied
in a perturbative sense, order by order. 
That is obviously true for the diverging parts which are the same as in the $\overline{MS}$ scheme 
and then satisfy Eq.(\ref{SMOM2}) at each order of the screened expansion.

By a dimensional argument, the finite parts of the self-energies can be written as power expansions
\begin{align}
\Pi_{loop}^{finite}(p)&=-p^2\left[\alpha \,F(s)+\alpha^2 F_2(s)+\cdots\right]\nn\\
\Sigma_{loop}^{finite}(p)&=p^2\left[\alpha \,G(s)+\alpha^2 G_2(s)+\cdots\right]\nn\\
\label{loops}
\end{align}
where $s=p^2/m^2$ and we use the short hand notation
\BE
\alpha=3N\left(\frac{\alpha_s}{4\pi}\right),\quad \alpha_s=\frac{g^2}{4\pi}.
\label{alpha}
\EE
The functions $F(s)$, $G(s)$, $F_2(s)$, etc., are adimensional functions of the variable $s$ and do not contain any parameter.
Explicit analytical expressions for the one-loop functions $F(s)$ and $G(s)$ were evaluated in Ref.\cite{ptqcd2} up to third order
in the $\delta$-expansion and are reported in Appendix A.

At one-loop, in the SMOM scheme, the subtracted self energies in Eq.(\ref{selfs}) can be written as
\begin{align}
\Pi(p)&=-p^2\left[\alpha\, F(s)-\frac{1}{s}-\alpha\, F(t)+\frac{1}{t}\right] \nn\\
\Sigma (p)&=p^2\left[\alpha\, G(s)-\alpha \,G(t)\right]
\label{selfsMOM}
\end{align}
where $t=\mu^2/m^2$ is the renormalization scale in units of $m$. In the above equation, the normalization
conditions of Eq.(\ref{SMOM1}) are evident since $\Pi(\mu)=\Sigma(\mu)=0$.
Moreover, any arbitrary additive constant in the definition of the functions $F(s)$, $G(s)$ is subtracted and
made irrelevant. By comparison with Eq.(\ref{selfs}), the finite parts of the renormalization constants follow
\begin{align}
\delta Z_A&=-\alpha\, F(t)+\frac{1}{t}\nn\\
\delta Z_c&=-\alpha\, G(t)
\label{deltaZ}
\end{align}
We observe that the two terms in $\delta Z_A$ are of different orders. The second one is of first order in the $\delta$-expansion (i.e counting
the number of vertices) but does not vanish in the limit $\alpha\to 0$ since it arises from the two-point vertex $\delta \Gamma^{\mu\nu}$ 
which adds a tree-level term $m^2$ to the gluon polarization in Eq.(\ref{selfs}). 

Inserting in Eq.(\ref{dressprop}) and using a screened definition of the gluon dressing function $J_m(s)$
\begin{align}
J_m(s)&=(p^2+m^2)\,\Delta (p)= p^2 \,\Delta(p)\, \left(1+\frac{1}{s}\right)\nn\\
\chi_m(s)&=-p^2\,{\cal G}(p)
\label{dressdef}
\end{align}
we can write the renormalized dressing functions at the scale $t$ as
\begin{align}
J_m(t,s)&=\frac{\displaystyle \left(1+\frac{1}{s}\right)}
{\displaystyle \left(1+\frac{1}{t}\right)+\alpha(t)\left[F(s)-F(t)\right]}\nn\\
\chi_m(t,s)&=\frac{1}{1+\alpha(t)\left[G(s)-G(t)\right]}
\label{dressing0}
\end{align}
where $\alpha (t)$ is the renormalized couplig $\alpha$ at the scale $t$.
It can be easily checked that $J_m(t,t)=\chi_m(t,t)=1$, as required by the SMOM scheme.

In the standard MOM scheme, the gluon dressing function is defined as
$J_0(s)=p^2\,\Delta(p)$ and by comparison with the definition of $J_m(s)$, 
Eq.(\ref{dressdef}), the two functions are related by
\BE
J_m(s)=J_0(s)\,\left(1+\frac{1}{s}\right).
\label{J0}
\EE

Multiplicative renormalization requires that we can define
renormalization constants such that the ratios
\begin{align}
\frac{Z_A(t^\prime)}{Z_A(t)}&=\frac{J_m(t,s)}{J_m(t^\prime, s)}=J_m(t,t^\prime)\nn\\
\frac{Z_c(t^\prime)}{Z_c(t)}&=\frac{\chi_m(t,s)}{\chi_m(t^\prime, s)}=\chi_m(t,t^\prime)
\label{ZAc}
\end{align}
do not depend on the momentum $s$. Here, the last terms of the equalities arise by setting $s=t^\prime$.
Quite generally, the one-loop dressing functions can be multiplicatively renormalized, but in a very limited
momentum range and for a very small coupling $\alpha\ll 1$. 
For instance, for the ghost case, in the limited range  $s\approx t\approx t^\prime\gg 1$ 
we can set $\alpha=\alpha(t)=\alpha(t^\prime)+{\cal O}(\alpha^2)$ and using Eq.(\ref{dressing0})
\begin{align}
\frac{Z_c(t^\prime)}{Z_c(t)}&=\frac{1+\alpha\left[G(s)-G(t^\prime)\right]}
{1+\alpha\left[G(s)-G(t)\right]}\approx\nn\\
\approx&\frac{1}{1+\alpha\left[G(t^\prime)-G(t)\right]}
=\chi_m(t,t^\prime)+{\cal O}(\alpha^2)
\label{chichain}
\end{align}
which does not depend on $s$ provided that $\alpha\left[G(s)-G(t)\right]\ll 1$.

That limited range can be extended numerically by evaluating RG improved dressing functions. 
Because of the special nature of the screened expansion, especially for the gluon dressing function, it
is instructive to recover the RG integration by a direct calculation rather than using the anomalous dimensions.
We will check that the same result is obtained by the anomalous dimensions if they are evaluated from the explicit 
one-loop renormalization constants of Eq.(\ref{deltaZ}).

Using their definition, Eq.(\ref{dressing0}), the one-loop gluon dressing functions satisfy
\begin{align}
J_m(t,t_0)^{-1}&\,J_m(t_0,t^\prime)^{-1}=\nn\\
=&1+\frac{\alpha(t)\left[F(t^\prime)-F(t)\right]-\left(\frac{1}{t^\prime}-\frac{1}{t}\right)}
{1+\frac{1}{t^\prime}}+{\cal O}(\delta t^2)\nn\\
&=J_m(t,t^\prime)+{\cal O}(\delta t^2)
\label{chain}
\end{align}
where $\delta t=t^\prime-t$. Then, in the limit $\delta t_i=(t_{i+1}-t_{i})\to 0$, setting $t_1=t$ and $t_{N+1}=s$,
\begin{align}
\ln J_m(t,s)^{-1}&=\lim_{\delta t_i\to 0}\sum_{i=1}^N \ln J_m(t_i,t_{i+1})^{-1}=\nn\\
=&\int_t^s\frac{\alpha(x) F^\prime (x)+x^{-2}}{1+x^{-1}}\,{\rm d}x
\label{RGint}
\end{align}
where $F^\prime (x)$ is the derivative of the function $F(x)$.
The second term can be integrated exactly yielding
\BE
J_m(t,s)=\left(\frac{1+s^{-1}}{1+t^{-1}}\right)\,\exp\left[{-\int_t^s\frac{\alpha(x) F^\prime(x)}{1+x^{-1}}\, {\rm d}x}\right].
\label{JRG}
\EE
For the ghost dressing function, from Eq.(\ref{chichain}) the same calculation gives
\BE
\chi_m(t,s)=\exp\left[-\int_t^s \alpha(x) G^\prime(x)\, {\rm d}x\right].
\label{chiRG}
\EE
We can easily check that,  since ${\rm d}\mu/\mu={\rm d} t/(2t)$, Eq.(\ref{RGint}) and Eq.(\ref{chiRG}) assume the
more familiar shape
\begin{align}
J_m(s_0,s)&=\exp\left[ \int_{s_0}^s \frac{{\rm d} t}{2t}  \gamma_A(t) \right]\nn\\
\chi_m(s_0,s)&=\exp\left[ \int_{s_0}^s \frac{{\rm d} t}{2t}  \gamma_c(t) \right]
\label{jchiRG}
\end{align}
where the anomalous dimensions $\gamma_A$, $\gamma_c$ are given by the same explicit expressions which follow, by a direct calculation,
from the renormalization constants of Eq.(\ref{deltaZ})
\begin{align}
\gamma_A(t)&=\mu \frac{\partial \ln(1+\delta Z_A)}{\partial\mu}=\frac{2t}{1+\delta Z_A}\frac{\partial }{\partial  t}\left[\frac{1}{t}-\alpha F(t)\right]\nn\\
=&-2t\frac{ \alpha\, F^\prime(t)+t^{-2}}{1+t^{-1}}+{\cal O}(\alpha^2)\nn\\
\gamma_c(t)&=\mu \frac{\partial \ln(1+\delta Z_c)}{\partial\mu}=\frac{2t}{1+\delta Z_c}\frac{\partial }{\partial  t}\left[-\alpha G(t)\right]\nn\\
=&-2t\,\alpha\, G^\prime(t)+{\cal O}(\alpha^2).
\label{gammas}
\end{align}
We notice that, at one loop, the tree-level term $1/t$ must be retained in the inverse of $(1+\delta Z_A)$ since it is not of order ${\cal O}(\alpha)$,
yielding the same identical result found in Eq.(\ref{RGint}) by a direct integration.

The running coupling $\alpha(t)$, which is required for the numerical integration in Eqs.(\ref{JRG}), (\ref{chiRG}), 
follows from ST identities which in the SMOM scheme, through Eqs.(\ref{SMOM2}),(\ref{ZAc}) yield\cite{alkofer}
\BE
\frac{\alpha(t)}{\alpha(t_0)}=\left[\frac{Z_g(t_0)}{Z_g(t)}\right]^2=J_m(t_0,t)\,\chi_m(t_0,t)^2
\label{SMOMrun1}
\EE
which completes the definition of the RG improved one-loop dressing functions. 
The closed set of non-linear coupled integral equations, Eqs.(\ref{JRG}), (\ref{chiRG}) and (\ref{SMOMrun1}) can be solved numerically, starting
from some initial point $\alpha(t_0)$.

We are not pursuing the approach further since we are mainly interested in an analytical description of the IR limit, say below 2 GeV, where the 
fixed-coupling optimized screened expansion already provides excellent results.\footnote{The present RG 
approach might not be reliable deep in the IR, 
since the scale-dependent anomalous dimensions in
Eq.(\ref{gammas}) have been derived by one-loop expressions 
which can only be trusted at an optimal scale $\mu\approx m$, as discussed in the next section.
While the one-loop anomalous dimensions are usually good enough, their validity in the present RG approach
has not been proven yet.}  
Actually, the main advantage of the fixed-coupling expansion is that
it provides simple analytical expressions which can be continued to the whole complex plane in order to explore the analytic properties. A feature which
makes the expansion a unique tool for accessing information that is not currently reachable by lattice calculations or other numerical methods.
Eventually, the RG improved dressing functions might be useful for a matching between the optimized expansion and ordinary PT above 2 GeV, since the
anomalous dimensions $\gamma_A$, $\gamma_c$ in Eq.(\ref{gammas}) tend to the standard result of ordinary PT in the limit $t\gg 1$ ($\mu\gg m$).

\section{Optimized Screened Expansion}

In the SMOM scheme, the  fixed-coupling one-loop dressing functions have explicit analytical expressions given by Eq.(\ref{dressing0}) where
the finite part of the self-energies is written in terms of the adimensional functions $F(s)$, $G(s)$ which are reported in Appendix A.

We can group together the constants in the denominators of Eq.(\ref{dressing0})
and define the scale-dependent constants
\begin{align}
F_0(t)=\left[\frac{1}{\alpha(t)}\right]\left(1+\frac{1}{t}\right)-F(t)\nn\\
G_0(t)=\left[\frac{1}{\alpha(t)}\right]-G(t).
\label{F0G0}
\end{align}
so that the dressing functions can be recast as
\begin{align}
J_m(t,s)&=\left[\frac{1}{\alpha(t)}\right]\frac{\displaystyle \left(1+\frac{1}{s}\right)}{F(s)+F_0(t)}\nn\\
\chi_m(t,s)&=\left[\frac{1}{\alpha(t)}\right]\frac{1}{G(s)+G_0(t)}.
\label{dressing}
\end{align}
For future reference, we also introduce two differently normalized dressing functions
\begin{align}
J(t,s)&=\alpha(t)\,J_m(t,s)=\left(1+\frac{1}{s}\right)\frac{1}{F(s)+F_0(t)}\nn\\
\chi(t,s)&=\alpha(t)\,\chi_m(t,s)=\frac{1}{G(s)+G_0(t)}
\label{dressing2}
\end{align}
which at $s=t$ satisfy
\BE
J(t,t)=\alpha(t)=\chi(t,t).
\label{jchi}
\EE

When recast as in Eq.(\ref{dressing}), the fixed-coupling dressing functions are known to approach the exact result if the constants
$F_0$, $G_0$ take an optimal value\cite{ptqcd2,scaling,xigauge}. Only above 2 GeV some deviations occur, as usual in PT, 
because of the large logarithms $\ln (p/m)$ which require a RG improvement of the expansion.
It is remarkable that the first derivatives of the inverse
dressing functions $J_0(s)^{-1}=(1+s^{-1})\,J(s)^{-1}$ and $\chi(s)^{-1}$, as defined in Eq.(\ref{dressing2}), do not depend on any parameter and
are fully determined, since the constants $F_0$, $G_0$ are canceled by the derivative.
Actually, below 2 GeV, all lattice data of different authors, for $N=2,3$, collapse on such curves when the energy units are properly set\cite{scaling}. 

By inspection of Eq.(\ref{dressing}), we observe that, up to an overall factor, the shape of the dressing
functions (as functions of the momentum $s$) depends on the scale-dependent constants $F_0(t)$, $G_0(t)$, which is at odds with the concept of
multiplicative renormalization. Moreover, since the constants depend on the subtraction point $t$,
they are rather arbitrary and, in principle, even different subtraction points could be taken for
the two functions. In previous works, the constants $F_0$ and $G_0$ have been used as independent variational parameters for optimizing the
one-loop expansion\cite{ptqcd2,scaling,xigauge}. The remarkable agreement with the lattice data says that
an optimal choice of the constants mimics the neglected higher order terms. Actually, at an optimal subtraction point
the higher-order  terms, which are also scale dependent, must be so small that the one loop calculation approaches the
exact result.

In the SMOM scheme, the constants $F_0$, $G_0$ are not independent since 
they are defined by Eq.(\ref{F0G0}) and depend on the scale.
It turns out that a very predictive self-contained variational criterion can be established for determining an optimal subtraction
point where the scale-dependent constants $F_0(t)$, $G_0(t)$ are stationary and the dressing functions 
approach the exact result. Thus, the optimization of the expansion can be reached  by first principles, without any input from the lattice.

If we trust Eq.(\ref{dressing}), it is obvious that the ratios $J_m(t,s)/J_m(t^\prime,s)$
and $\chi_m(t,s)/\chi_m(t^\prime,s)$
in Eq.(\ref{ZAc}) can be totally independent of the  momentum $s$ if and only if the constants $F_0$, $G_0$ are RG invariants i.e.
\BE
\frac{\partial G_0(t)}{\partial t}=\frac{\partial F_0(t)}{\partial t}=0
\label{RGinv}
\EE
which is generally not the case for the approximate one-loop expressions of $G_0$, $F_0$ in Eq.(\ref{F0G0}). 
Thus, if there is an optimal subtraction point $t=t^\star$ where the one-loop dressing functions $J_m(t^\star,s)$, $\chi_m(t^\star,s)$ approach
the exact result, then the functions $G_0$, $F_0$ should be at least stationary at $t=t^\star$, so that the shape of the dressing functions
in Eq.(\ref{dressing}) does not change for a small shift of the scale around $t^\star$, as required by the multiplicative renormalization of
the exact dressing functions. In other words, the optimal scale $t^\star$ must be the solution of Eq.(\ref{RGinv}), which is the 
point where the constants $G_0$, $F_0$ are locally RG invariant. We expect that any prediction of the one-loop expressions can only be reliable
at $t=t^\star$ where $G_0(t^\star)$, $F_0(t^\star)$ and $\alpha(t^\star)$ get closer to their exact values. An other, related, point of view is
that since the shape of the one-loop dressing functions depend on the constants $F_0$, $G_0$ while the shape  of the exact dressing functions 
must be RG invariant, then the best subtraction point must be where the sensitivity to any scale change is minimal, according to Stevenson's principle
of minimal sensitivity\cite{sensitivity,stevensonRS}. Again, we find the stationary condition of Eq.(\ref{RGinv}).

The nice thing is that, once we have got a good approximation for the dressing functions and the coupling at $t=t^\star$, we can easily extend
the result at any scale $t$ by the exact scaling equations of the SMOM scheme, Eqs.(\ref{SMOM2}),(\ref{ZAc}), without having to rely on
the approximate one-loop relations which are very poor away from $t=t^\star$. From Eq.(\ref{ZAc}) we can write
\begin{align}
J_m(t,s)&=\frac{J_m(t^\star,s)}{J_m(t^\star, t)},\quad
\chi_m(t,s)=\frac{\chi_m(t^\star,s)}{\chi_m(t^\star, t)}
\label{JchiRen}
\end{align}
while from ST identities, Eq.(\ref{SMOMrun1}) reads
\BE
\alpha_T(t)=\alpha_T(t^\star)\,J_m(t^\star,t)\,\chi_m(t^\star,t)^2
\label{SMOMrun}
\EE
yielding an explicit analytical expression for the running coupling $\alpha_T$ in the Taylor-SMOM scheme.

In more detail, we observe that the definition of $F_0$, $G_0$, in Eq.(\ref{F0G0}), 
arises from the one-loop approximation and is biased by the truncation of the expansion. 
While we expect the error to be minimal or vanishing at the optimal scale $t=t^\star$, 
where the higher order corrections might cancel, in general Eq.(\ref{F0G0}) is not reliable away from $t=t^\star$. 
Thus, if we replace $F_0$, $G_0$ by the correct constant values that make the dressing functions multiplicatively 
renormalizable, then the coupling $\alpha(t)$ in Eq.(\ref{F0G0}) cannot
be equal to the correct coupling $\alpha_T(t)$ which arises from the exact Eq.(\ref{SMOMrun}), 
but it would provide a poor approximation away from $t^\star$.
The one-loop function $\alpha(t)$ might be regarded as an effective coupling 
incorporating the neglected higher order terms and such that $\alpha(t^\star)=\alpha_T(t^\star)$. 
As observed above, no such effective coupling can exist if $F_0$ and $G_0$ are constants: 
Eq.(\ref{F0G0}) can be inverted as a definition of $\alpha$ and the two emerging equations would be incompatible,
for any choice of the constants. 
However, if we just impose that the RG invariance condition, Eq.(\ref{RGinv}), is satisfied at a single point, 
$t=t^\star$, then a solution is found and Eq.(\ref{RGinv}) can be regarded as a stationary condition 
for determining the optimal scale $t^\star$. 

\begin{figure*}[htp] \label{fig:solution}
\centering
\includegraphics[width=0.32\textwidth,angle=-90]{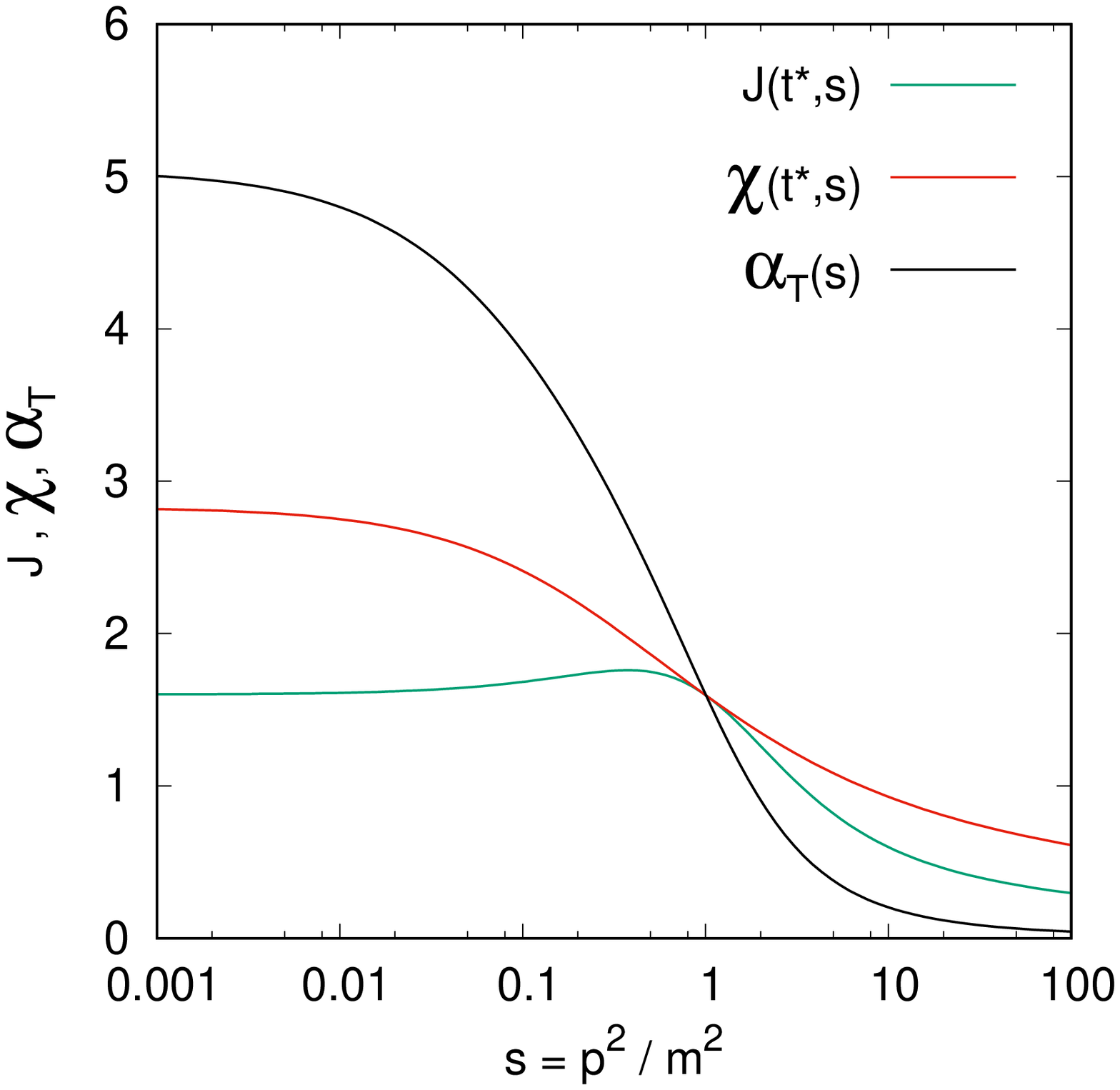}\includegraphics[width=0.32\textwidth,angle=-90]{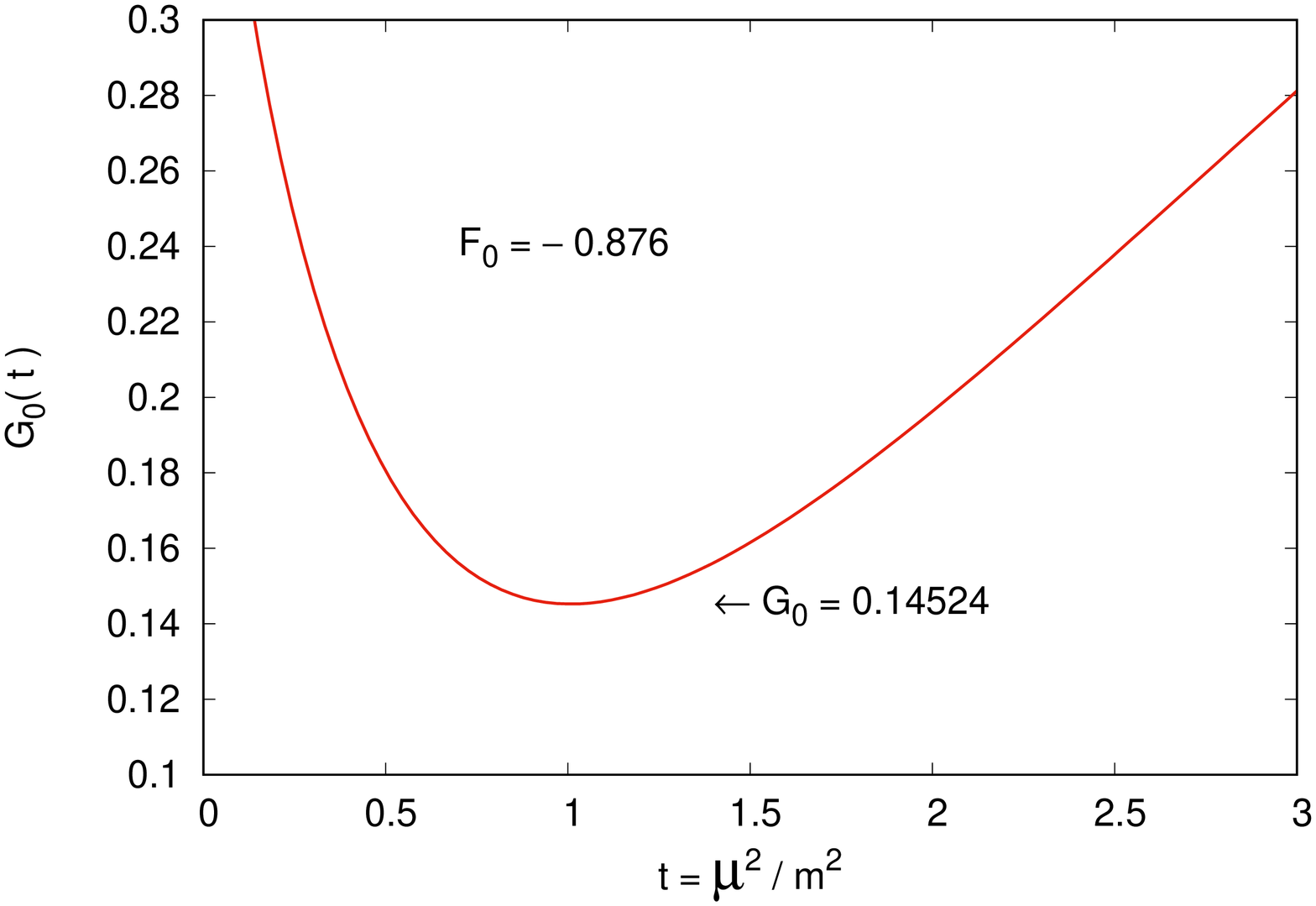}
\caption{Graphical solution of Eq.(\ref{crossing}). In the left panel,
the dressing functions $J(t^\star,s)$, $\chi(t^\star, s)$ of Eq.(\ref{dressing2}) are shown together with the SMOM running coupling
$\alpha_T(s)$ of Eq.(\ref{SMOMrun}) for the optimal set $F_0(t^\star)=-0.876$, $G_0(t^\star)=0.14524$, 
$t^\star=1.00749$ and $\alpha_T(t^\star)=\alpha(t^\star)=1.5939$. 
In the right panel, the same solution appears as the 
stationary point of $G_0(t)$ when $F_0=-0.876$ in Eq.(\ref{G0}).}
\end{figure*}

In fact,
the stationary condition can be recast as the solution of two coupled equations
for the dressing functions $J$, $\chi$, as defined in Eq.(\ref{dressing2}). Using Eq.(\ref{jchi}), we 
observe that, whatever is the unknown function $\alpha$, 
the total derivative of $J(t,t)$ and $\chi(t,t)$ must coincide, so that if the stationary condition,
Eq.(\ref{RGinv}), is satisfied at $t=t^\star$, then
\begin{align}
\frac{\partial J(t^\star,s)}{\partial s}\Big\rvert_{s=t^\star}&=\frac{\partial \chi(t^\star,s)}{\partial s}\Big\rvert_{s=t^\star}\nn\\
J(t^\star,s)\Big\rvert_{s=t^\star}&=\chi(t^\star,s)\Big\rvert_{s=t^\star}=\alpha(t^\star)
\label{crossing}
\end{align}
and the scale $t^\star$ emerges as the point of tangency of the dressing functions $J(t^\star,s)$, $\chi(t^\star,s)$, while
$\alpha(t^\star)=\alpha_T(t^\star)$ is the actual value of the functions at the crossing point.

The three conditions in Eq.(\ref{crossing}) are not enough for determining the four 
unknown variables $t^\star$, $F_0(t^\star)$, $G_0(t^\star)$
and $\alpha(t^\star)$, but they provide a link among them and all of them are fully determined if one is known.
We observe that the actual value of the constants $F_0$, $G_0$ depends on the definition of the functions $F(s)$, $G(s)$, since adding a constant
to $F$  (or $G$) and subtracting it from $F_0$ (or $G_0$), leaves Eqs.(\ref{dressing}) and (\ref{F0G0}) unchanged. 

Using the same definitions of Ref.\cite{ptqcd2} for these functions, as reported in Appendix A, the optimal value $F_0=-0.876$ was determined in Ref.\cite{xigauge} by
requiring that the complex poles and the whole principal part of the gluon propagator 
are gauge parameter independent,\footnote{Actually, in Ref.\cite{xigauge}, only the gauge-parameter independence
of the poles and of the phases of the residues was enforced, since they are RG invariant. The modulus of the bare residue is not a well-defined quantity because it diverges and needs renormalization.}
a feature that
the exact gluon propagator inherits from BRST symmetry through the Nielsen identities. For that value, without any fit of parameters, 
Eqs.(\ref{dressdef}) and (\ref{dressing}) provide a very accurate reproduction of the lattice data of Ref.\cite{duarte} if the data are scaled
in units of $m=0.656~{\rm GeV}$\cite{xigauge}.  
Since all predictions of the optimized screened expansion are in units of $m$, taking that value of $m$ is equivalent to 
share the same energy units of the data sets in Ref.\cite{duarte}.

The ghost dressing function has no poles and no information can be gained by Nielsen identities on the optimal value of $G_0$. However, by the
stationary condition of Eq.(\ref{crossing}), having set $F_0(t^\star)=-0.876$ by the analytic properties of the gluon propagator\cite{xigauge}, 
we obtain the optimal set of
variables $G_0(t^\star)=0.14524$, $t^\star=1.00749$ and $\alpha(t^\star)=1.5939$ ($\alpha_s=4\pi\alpha/9=2.2255$ for $N=3$). 

A graphical solution of the equations is shown in the left panel of Fig.~1 where
the functions $J(t^\star,s)$, $\chi(t^\star,s)$ are plotted together with the running coupling $\alpha_T$
of Eq.(\ref{SMOMrun}). As required by
Eq.(\ref{crossing}), the curves cross at the same point where the dressing functions are also tangent. 
We observe that the running coupling that arises in the SMOM scheme is monotonic, with a finite IR stable fixed point, 
at variance with the MOM scheme.
It is remarkable that the optimal subtraction point is so close to the mass parameter, $\mu^\star=\sqrt{t^\star}\>m=1.004\, m$. It is not 
totally unexpected
since terms like $\ln(\mu/m)$ are minimized at that scale.

While no special assumption has been made for the functions $F_0(t)$, $G_0(t)$, 
besides their stationarity at $t=t^\star$, 
it is instructive to see how the same solution is found at $t^\star$ if the function $F_0(t)$ is assumed to be strictly
constant and equal to the same optimal value $F_0(t)=F_0(t^\star)=-0.876$ which was dictated by Nielsen identities. 
With that choice, the poles are gauge parameter independent and the multiplicative 
renormalization of the gluon dressing function is explicitly satisfied in Eq.(\ref{dressing}). However, 
the exact multiplicative renormalization of the ghost dressing function cannot be recovered because the function
$G_0(t)$ is not a constant according to Eq.(\ref{F0G0}) which can be recast, by eliminating the coupling, as
\BE
G_0(t)=\left[F_0+F(t)\right]\left(\frac{t}{1+t}\right)-G(t)
\label{G0}
\EE
and is plotted in the right panel of Fig.~1. A stationary point occurs at $t=t^\star$, which emerges as the optimal
scale where the multiplicative renormalization of the ghost dressing function is approximately recovered.

Finally, we emphasize that the one-loop predictions might be very poor away from $t=t^\star$. Even the effective
coupling $\alpha(t)$, which will be shown to be very close to the exact value at $t=t^\star$, is not reliable
at a generic scale $t$. 
In fact, Fig.~1 already shows that its derivative, which gives the slope of the 
dressing functions according to Eqs.(\ref{jchi}),(\ref{RGinv}), is just one third of the correct slope of $\alpha_T$,
as arising from the exact Eq.(\ref{SMOMrun}). In more detail, at the stationary point $t^\star$, because of the vanishing of the derivatives in Eq.(\ref{RGinv}), the derivative of the effective coupling is
\BE
\alpha^\prime(t^\star)=
\frac{\partial J(t^\star,s)}{\partial s}\Big\rvert_{s=t^\star}=\frac{\partial \chi(t^\star,s)}{\partial s}\Big\rvert_{s=t^\star}
\EE
and the derivative of the running coupling $\alpha_T$ follows from Eq.(\ref{SMOMrun}) using 
Eqs.(\ref{crossing}),(\ref{dressing2})
\BE
\alpha_T^\prime(t^\star)=\alpha_T(t^\star)\frac{3\alpha^\prime(t^\star)\left[\alpha(t^\star)\right]^2}
{\left[\alpha(t^\star)\right]^3}=3\,\alpha^\prime(t^\star).
\EE
Strictly speaking, because of the vanishing of the derivatives in Eq.(\ref{RGinv}), without any further assumption on
the functions $F_0(t)$, $G_0(t)$, the optimization conditions give an equation for a fifth variable,
the derivative $\alpha^\prime(t^\star)$, which is fully determined, but is one third of the correct value.
Thus, not all the one-loop predictions approach the correct value at the stationary point. 
Even if higher order terms somehow conspire to cancel in the dressing functions at $t=t^\star$, 
the same feature is not guaranteed to occur in the calculation of all quantities.

\section{Predictions of the Optimized Expansion}

Having found the optimal subtraction point $\mu^2/m^2=t^\star$ for the fixed-coupling screened expansion, we are left with explicit analytical
expressions for the propagators and the strong coupling which are fully determined, without any free parameter, from first principles. 
In this section, the predictions of the optimized expansion are compared with the numerical result of some recent lattice calculations.

In Fig.2 the ghost dressing function $\chi_m(t^\star,s)$ and the gluon propagator $m^2 \Delta(p)=J_m(t^\star,s)/(1+s)$, renormalized at
the optimal scale $t^\star$, are shown together with the large-volume lattice data of Ref.\cite{duarte} ($\beta=6.0$, $L=80$). The lattice
data have been
multiplicatively renormalized at the scale $t^\star\approx 1$ and reported in adimensional units by taking $m=0.656$ GeV as in Ref.\cite{xigauge}.
The plots are not a fit of the data since everything is determined apart from the energy scale $m$ which 
is obviously dictated by the data.
The perfect agreement with the data emerges from the stationary condition, Eq.(\ref{RGinv}), and from the gauge-parameter independence of
the poles, as required by ST identities and Nielsen identities, respectively. All such features are a consequence of the exact BRST symmetry 
of the total action. Since the symmetry is broken at any finite order of the screened expansion, the one-loop dressing functions give a better
approximation at the optimal scale $t=t^\star$ where  the identities are best satisfied.

\begin{figure*}[htp] \label{fig:Jchi}
  \centering
  \includegraphics[width=0.32\textwidth,angle=-90]{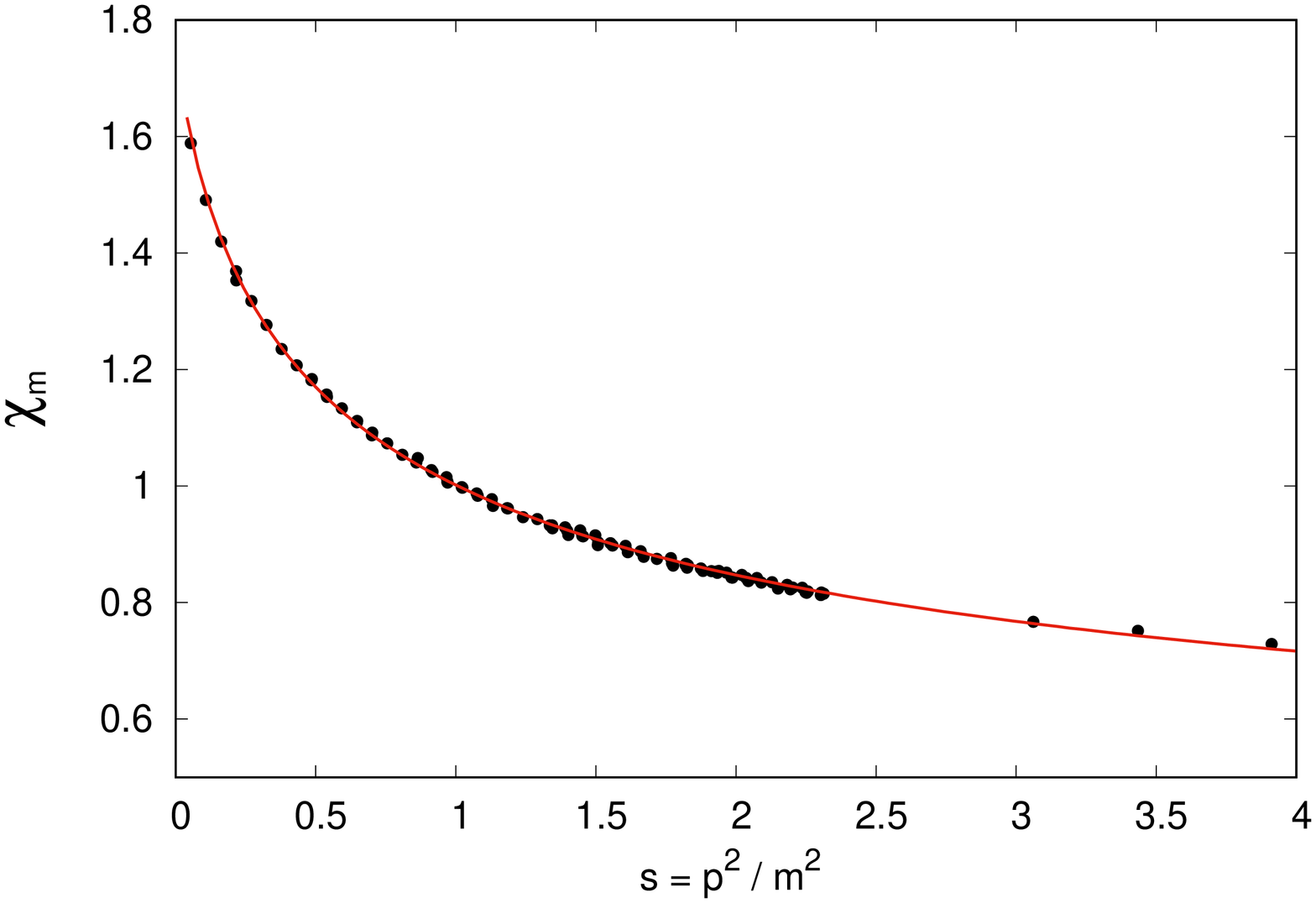}\quad\includegraphics[width=0.32\textwidth,angle=-90]{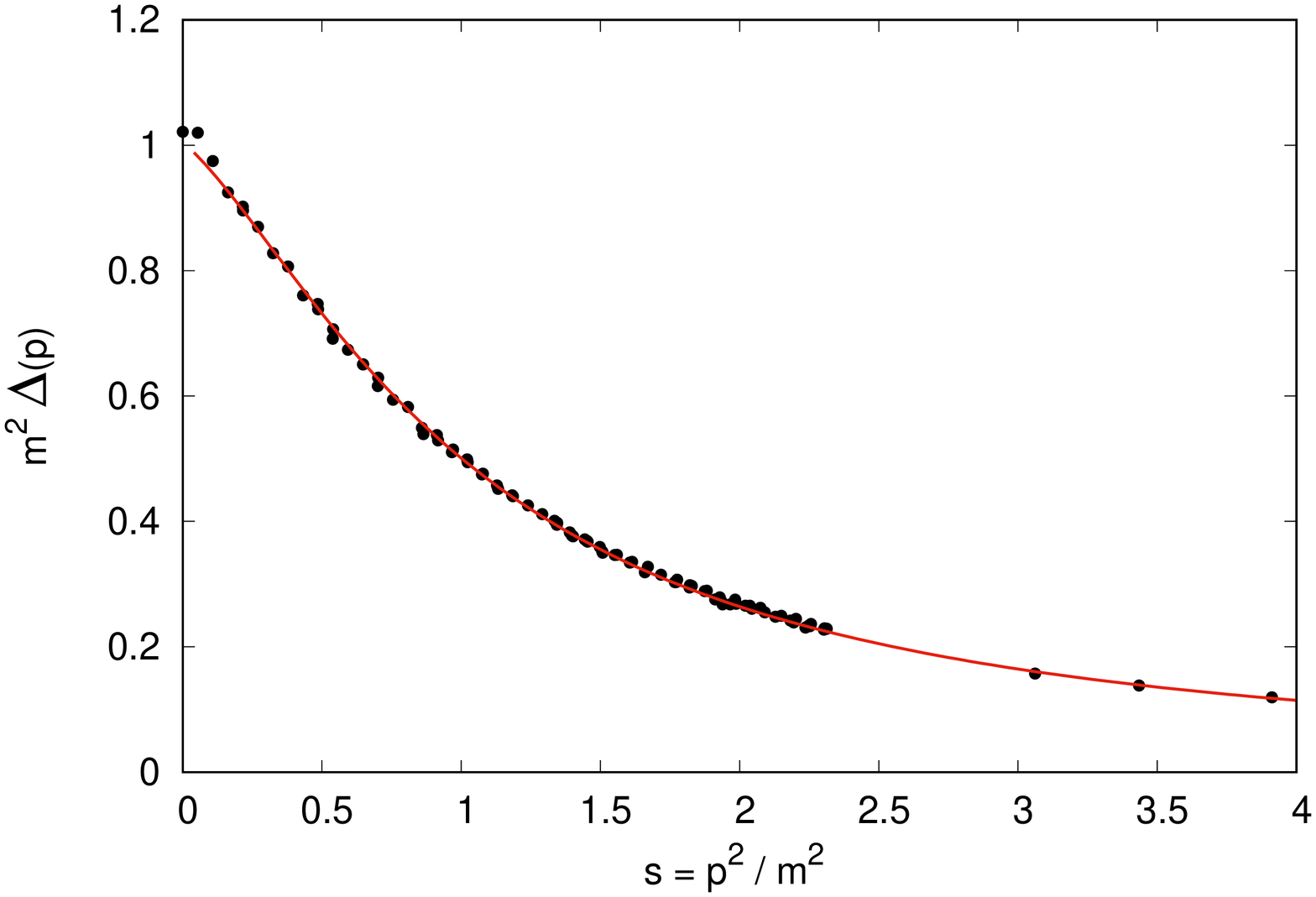}
\caption{The ghost dressing function $\chi_m(t^\star, s)$ (left) and the gluon propagator $m^2\,(1+s)\,J_m(t^\star,s)$ (right) are
evaluated by Eq.(\ref{dressing}) and shown together with the lattice data of Ref.\cite{duarte}
for $\beta=6.0$ and $L=80$. The functions are renormalized at the optimal scale $t^\star=1.00749$ where $F_0(t^\star)=-0.876$ and
$G_0(t^\star)=0.14524$.
The lattice data are reported in units of $m=0.656$ GeV and renormalized at the scale $t^\star\approx 1$ by the factors $Z_\chi=0.572$, 
$Z_J=0.237$, such that $Z_\chi\,\left[\chi_{data}(t^\star)\right]=1$  and $Z_J\,m^2\,(1+t^\star)\,\left[\Delta_{data}(t^\star)\right]=1$.}     
\end{figure*}

\begin{figure*}[htp] \label{fig:alpha}
  \centering
\hspace*{-1cm}\includegraphics[width=0.45\textwidth,angle=-90]{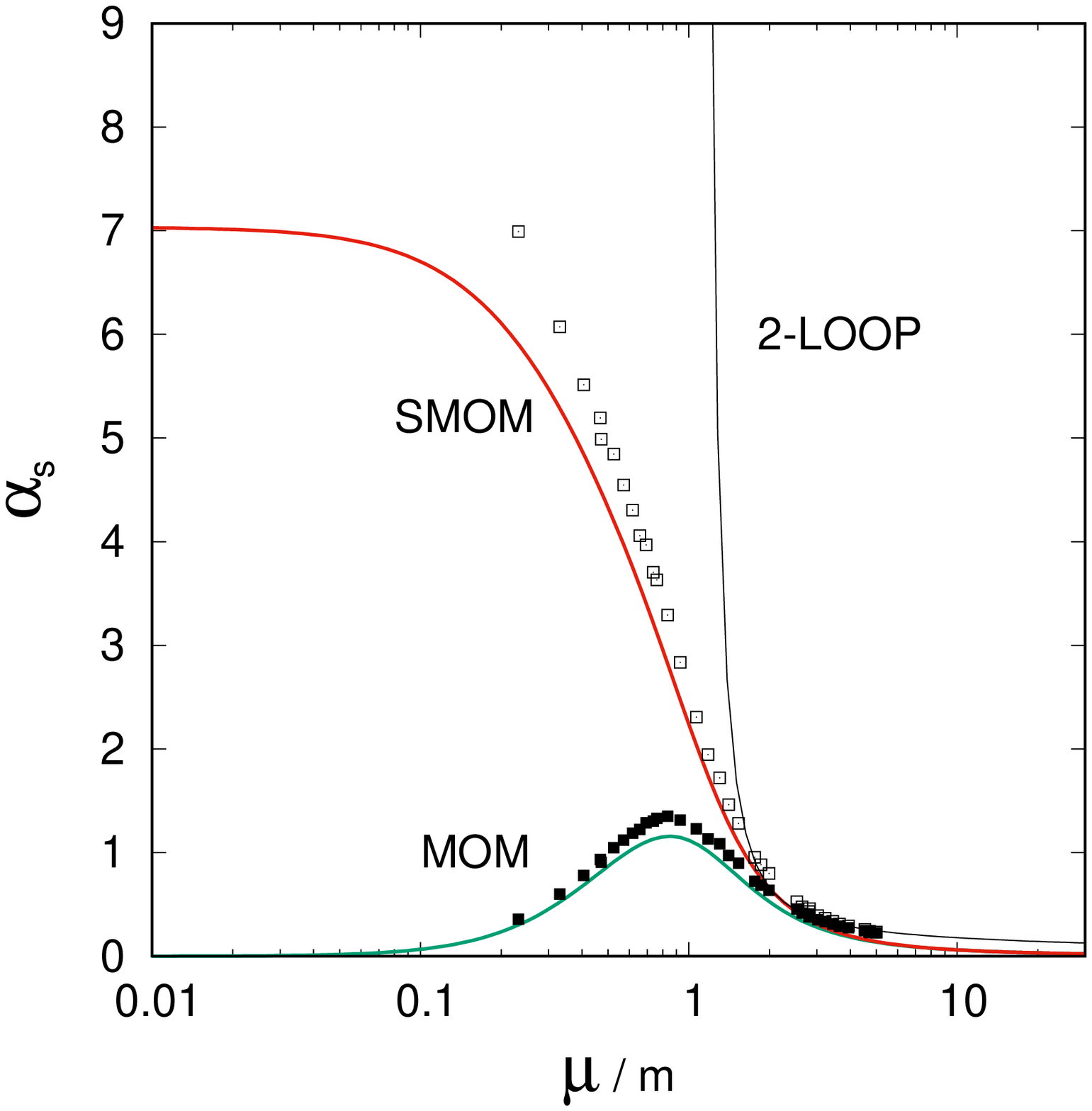}\hspace*{-3cm}\includegraphics[width=0.45\textwidth,angle=-90]{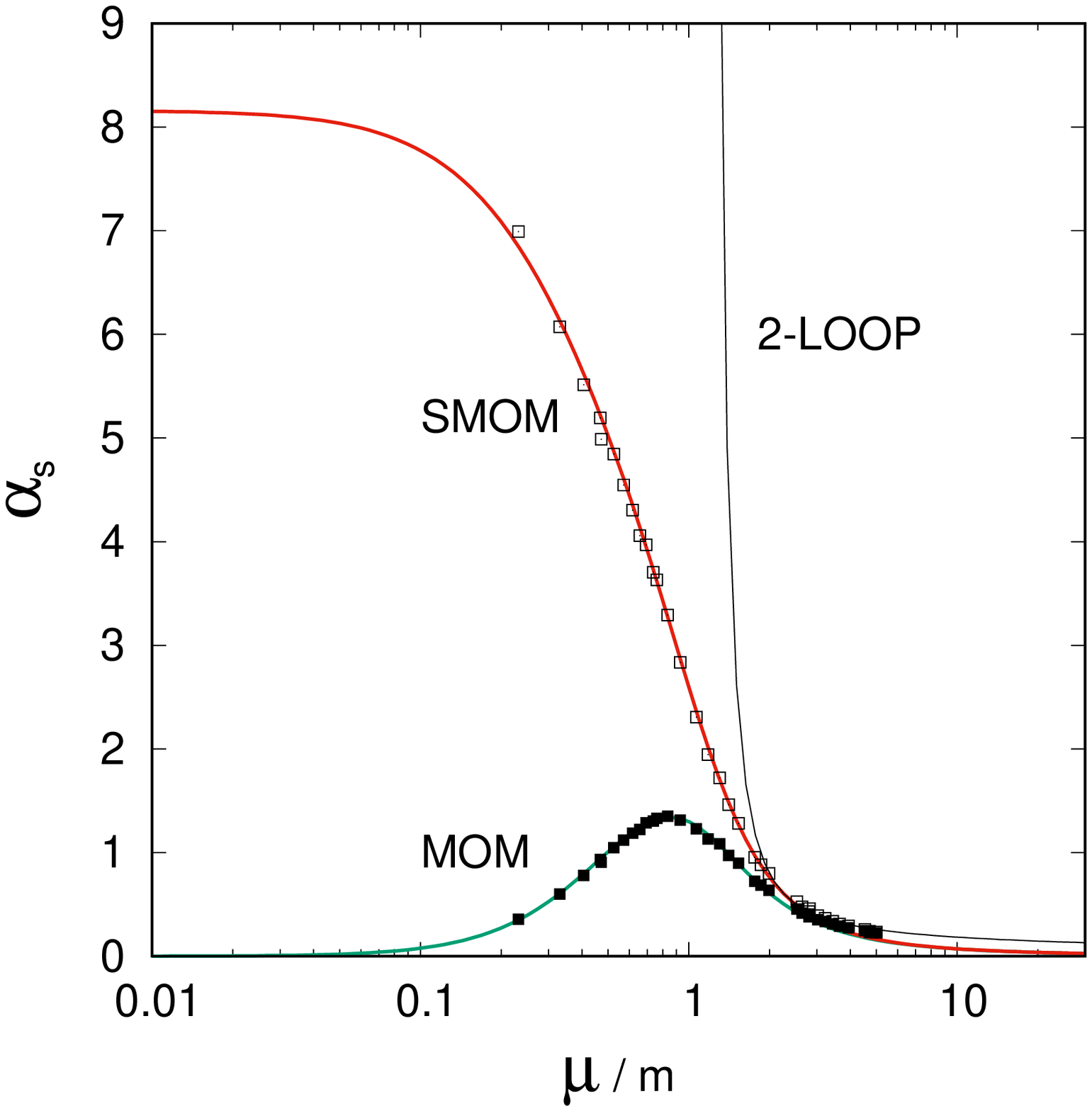}
\caption{In the left panel, the strong coupling $\alpha_s$ in the SMOM scheme (red line) 
and in the MOM scheme (green line) are shown together with the
lattice data of Ref.\cite{duarte} for $\beta=6.0$ and $L=80$ in the MOM scheme, in units of $m=0.656$ GeV (filled squares). 
The coupling is evaluated in the SMOM scheme by Eq.(\ref{SMOMrun}), 
using the optimal value $\alpha_s(t^\star)=2.2255$ predicted by the stationary conditions for $N=3$, 
and then rescaled in the MOM scheme by Eq.(\ref{scaling}). The SMOM data
(open squares) are also rescaled  from the MOM data by Eq.(\ref{scaling}). In the right panel, 
the same curves are multiplied by a factor 1.16 in order to match the data. The standard two-loop result of Eq.(\ref{approx}) is also shown
for $\Lambda=0.999\, m$ (left panel) and $\Lambda=1.08\, m$ (right panel).}     
\end{figure*}

The prediction of the optimal constant $G_0$, from first principles, is encouraging and gives us confidence in the validity of the SMOM scheme
for the screened expansion. In fact, if the dressing function were renormalized in the standard MOM scheme,  the curves $J_0$, $\chi$ would not
even cross in Fig.~1. The ghost dressing function plays an important role for determining the running coupling by Eq.(\ref{SMOMrun}) and the
good agreement in Fig.~2 says that the running must be correctly described, at least up to 2 GeV. Beyond that point, the fixed-coupling 
ghost dressing function is slightly suppressed and some RG improvement is required.

There is an important further prediction that arises from the graphical solution of Eq.(\ref{crossing}) in Fig.~1. At the optimal scale $t^\star$
the coupling is predicted to be $\alpha(t^\star)=1.5939$ which is $\alpha_s=2.2255$ for $N=3$ at $\mu=\mu^\star\approx m$. 
It means that the method is capable of {\it measuring} the coupling once the energy units are fixed, as it happens on the lattice. 
Reversing the argument, if the coupling is known at a given scale, then we get an independent {\it measure} of the energy units and $m$.
Thus, it is quite interesting to compare these measures with the lattice data. We observe that, because of Eq.(\ref{J0}) and Eq.(\ref{SMOMrun1}),
we can easily switch from the SMOM to the MOM scheme 
\BE
\left[\alpha (t)\right]_{SMOM}=\left(1+\frac{1}{t}\right)\,\left[\alpha (t)\right]_{MOM}
\label{scaling}
\EE
so that $[\alpha_s(m)]_{MOM}=0.5\,[\alpha_s(m)]_{SMOM}\approx 1.11$.

The strong coupling $\alpha_s$, evaluated by Eq.(\ref{SMOMrun}) using the predicted value $\alpha_s(t^\star)=2.2255$ for $N=3$, is shown 
in Fig.~3 together with the lattice data of Ref.\cite{duarte} for $\beta=6.0$ and $L=80$ in the MOM scheme. The data are shown 
in units of $m=0.656$ GeV, as in Fig.~2. In the same figure, using Eq.(\ref{scaling}), the calculated strong coupling is also shown in the MOM
scheme while the lattice data, which are in the MOM scheme, are also converted and shown in the SMOM scheme. As expected from the agreement in Fig.~2,
when rescaled in the MOM scheme, the calculated strong coupling has the same shape of the data, with a maximum occurring 
at the same point $\mu/m=0.85$ ($\mu=0.56$ GeV). However, the estimate of the lattice is $16\%$ higher than the numerical value 
predicted by the stationary conditions of Eq.(\ref{crossing}). As shown in the right panel of Fig.~3, 
if multiplied by a factor $1.16$ the curves collapse on the data. 
There are good reasons to believe that the disagreement might be caused by the finite lattice spacing which is $\approx 0.1$ fm for the data set we are
considering ($\beta=6$). In fact, as discussed by the same authors of Ref.\cite{duarte}, the numerical value of the coupling depends 
on the lattice spacing and seems to be overestimated on the lattice, since it decreases with the decrasing of the lattice spacing. Actually,
differences larger than $10\%$ have been reported for different lattice spacings. Thus, the prediction of the present calculation is probably
closer to the exact result than the lattice estimates. At the maximum, the MOM strong coupling is predicted to be $\alpha_s=1.157$. 

For $\mu> 3 m\approx 2$ GeV, the fixed-coupling ghost dressing function is suppressed because of the large logs. In fact, using the explicit
expressions of  Appendix A, for large values of $s$, $t$ the function $G(x)$ has the leading behavior 
$G(s)-G(t)\approx \ln(s/t)/4$ and the approximate multiplicative renormalization in Eq.(\ref{chichain}) is valid for $t=t^\star\approx 1$ only if
$\alpha \ln s/4\ll 1$ which is $\ln s\ll 2.5$ for $\alpha(t^\star)\approx 1.6$, yielding $p\ll 3.5\, m\approx 2.3$ GeV.
The problem might be cured by the RG equations, as discussed in Section III. In Fig.~3 we observe a suppression of the running coupling
which goes below the data for $\mu> 3\, m$, even when rescaled, as in the right panel. 
The predictions of the fixed-coupling expansion are not reliable beyond that point. We can attempt a matching with the standard running
coupling of PT which for a large momentum is very well described by the scheme-independent two-loop beta function
\BE
\beta(g)=-\beta_0\frac{g^3}{16\pi^2}-\beta_1\frac{g^5}{128 \pi^4}
\label{beta}
\EE
and has the approximate solution\cite{beta} in powers of $1/\ln\mu^2$
\BE
\alpha_s\approx\frac{4\pi}{\beta_0\ln(\mu^2/\Lambda^2)}\left[1-\frac{2\beta_1}{\beta_0^2}\frac{\ln\left((\ln(\mu^2/\Lambda^2)\right)}{\ln(\mu^2/\Lambda^2)}
\right]
\label{approx}
\EE
where, for pure Yang-Mills $SU(3)$ theory, $\beta_0=11$ and $\beta_1=51$.
As shown in Fig.~3, the approximate two-loop coupling is tangent to the SMOM curve at $\mu=2.11\, m$ if $\Lambda=0.999\, m$. At that point the
strong coupling is  $\alpha_s=0.59$ ($g=2.723$). If the curve is rescaled in order to match the data point (right panel), the tangency
is found for $\Lambda=1.08$ at $\mu=2.31\, m$ where $\alpha_s=0.58$. Of course, the matching cannot be trusted too much because it occurs at a
borderline $\mu\approx 2\, m$ which is too large for the fixed-coupling expansion and two small for the two-loop approximation of standard PT. While the data
seem to be well described by the approximate two-loop expression of Eq.(\ref{approx}) for $\mu> 4\, m$ the SMOM curve gives a reliable prediction
for $\mu< 2\, m$, leaving a narrow window where neither of them can be trusted. That said, it is remarkable that, since $\Lambda\approx m$, the Landau
pole is replaced by a mass parameter which gives a scale to the theory and seems to play the same role of $\Lambda$.  

\begin{figure}[b] \label{fig:beta}
\includegraphics[width=0.35\textwidth,angle=-90]{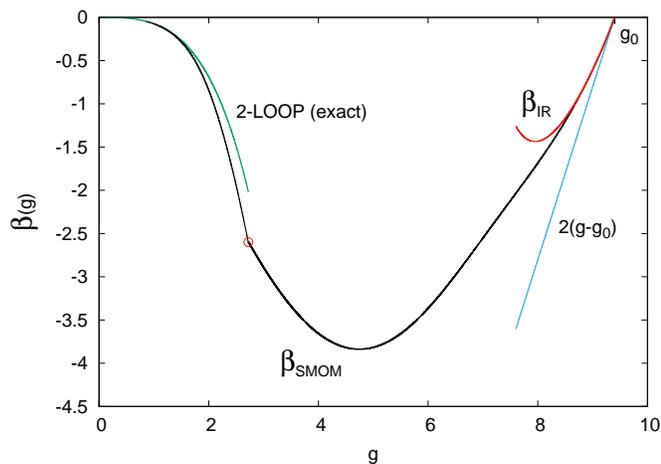}
\caption{The beta function in the SMOM scheme is obtained by the derivative of Eq.(\ref{SMOMrun}) 
and displayed as a black line for $g>2.723$ ($\mu <2.11\,m$) together with the outcome of the approximate two-loop result of Eq.(\ref{approx})
which is shown for $g<2.723$. The red circle is the matching point of the two curves at $g=2.723$. The beta meets the zero at $g=g_0\approx 9.40$.
The leading IR behavior of Eq.(\ref{beta2}), valid in the limit $g\to g_0$, is shown as a red line. 
The exact two-loop beta function of Eq.(\ref{beta}), which holds in the UV limit $g\to 0$, is shown as a green line.}
\end{figure}

In the IR, the SMOM scheme seems to be more appealing than the MOM scheme, since the strong coupling  saturates instead of decreasing, as one
would expect for a phenomenological coupling.
As shown in Fig.~3, at variance with the standard MOM scheme, in the screened scheme the strong coupling 
is a monotonic function of the momentum and reaches its maximum
at $\mu=0$ where $\alpha_s\approx 7.03$, with a vanishing derivative. Thus, the beta function meets a zero at $g=g_0\approx 9.40$
which is a finite IR-stable  fixed point. 

Except for the matching range at $\mu/m\approx 2-3$, Eq.(\ref{SMOMrun}) and Eq.(\ref{approx}) provide an explicit analytical expression for the strong
coupling and the beta function. No divergence occurs in the IR where no RG improvement seems to be mandatory since the mass parameter screens the theory
for $\mu\ll m$. 
The beta function is defined as
\BE
\beta(g)=\mu\,\frac{{\rm d} g}{{\rm d} \mu}=\left[2x\,\frac{{\rm d} g}{{\rm d} x}\right]_{x=x(g)}
\EE
and can be evaluated analytically by a simple derivative. 

In the SMOM scheme, the beta is obtained by Eq.(\ref{SMOMrun}) 
and is shown in Fig.~4 for $g>2.723$ ($\mu <2.11\,m$) together with the outcome of the approximate two-loop result of Eq.(\ref{approx})
which is shown for $g<2.723$. At the matching point, shown as a red circle at $g=2.723$, the curves tend to the same value because of
the tangency of the couplings in Fig.~3, but a jump in the derivative of the beta is observed. In fact, as discussed before, 
both curves are not reliable around the matching point and the exact beta function is expected to interpolate smoothly
between the two of them.
For comparison, in the UV limit $g\to 0$, the exact two-loop beta function of Eq.(\ref{beta}) is also plotted in Fig.~4, showing that the 
deviation of the approximate Eq.(\ref{approx}) is not small for $g>2$. 

As shown in Fig.~2 and Fig.~3, the optimized expansion is quite
reliable above the matching point, say for $g>3$, where the beta reaches a minimum at $g\approx 4.75$ and then increases and crosses
the zero at the IR stable fixed point $g_0\approx 9.40$. Some numerical values of the beta and of the coupling at specific point locations
are listed in Table I. 

\begin{table}[ht]
\centering 
\begin{tabular}{c c c c c c} 
\hline\hline 
Point location & $\mu/m$ & $g$ & $\beta$ & $\alpha_s$ & $\alpha_s$(MOM) 
\\ [0.5ex] 
\hline 
UV                        & $\infty$ &  0      &  0       &   0     & 0     \\ 
Matching point            & 2.11     & 2.723   &  -2.6    &  0.59   & 0.48  \\
Max of $\vert \beta\vert$ & 1.1575   & 4.745   &  -3.838  &  1.792  & 1.026 \\
Optimal scale $ t^\star$  & 1.0037   & 5.2883  &  -3.7437 &  2.2255 & 1.117 \\
Max of $\alpha_s$(MOM)    & 0.849    & 5.8915  &  -3.4386 &  2.762  & 1.157 \\
IR                        & 0        & 9.4017  &   0      &  7.034  & 0     \\
\hline 
\end{tabular}
\label{tableI} 
\caption{Values of the beta function and coupling at specific point locations, in the SMOM scheme, by
the screened expansion optimized at $t^\star=1.00749$ where $F_0(t^\star)=-0.876$ and
$G_0(t^\star)=0.14524$. In the last column, the strong coupling in the MOM scheme is also shown, converted by
Eq.(\ref{scaling}).}
\end{table}

In Appendix A, explicit expressions are derived for the leading behavior of the functions $F(x)$ and $G(x)$ in the limit
$x\to 0$. Deep in the IR, using Eq.(\ref{betaexpl}), the following explicit expression can be written for the leading behavior of the beta function 
\begin{align}
&\beta_{IR}(g)\approx 2(g-g_0)\times\nn\\
&\quad\times\left\{1-\left[\frac{a}{b}+\ln\frac{b\,g_0}{g-g_0}+\ln\left(\ln\frac{b\,g_0}{g-g_0}\right)\right]^{-1}\right\}
\label{beta2}
\end{align}

where  $g_0$ is the IR stable fixed point which can be exactly located as
\BE
g_0=\frac{4\pi}{\sqrt{3\,c\,N}}\>\frac{1}{k_0\,\alpha(t^\star)}\approx 9.4017 
\label{g00}
\EE
for $\alpha(t^\star)=1.5939$ and $N=3$,
while $a$, $b$, $c$ are numerical constants, defined in Appendix A and taking
the values
\BE
a\approx -0.3795,\quad b\approx -0.449, \quad c=5/8.
\EE

The leading behavior of Eq.(\ref{beta2}) is also shown in Fig.~4 in the IR limit $g\to g_0$
where it gives a very good approximation of the SMOM beta function for $g>8.5$. Asymptotically,
the beta tends to the linear behavior $\beta(g)=2(g-g_0)$.

\section{Discussion}

In the IR, the fixed-coupling one-loop screened expansion provides a complete and analytical
description of pure Yang-Mills theory if the exact constraints inherited by the BRST symmetry 
are imposed on the approximate one-loop expressions. While the symmetry is softly broken in the
expansion at any finite order, the total action has not been modified, so that the Nielsen and
ST identities must be still satisfied by the exact correlation functions of the theory.
On the other hand, since the same identities are not satisfied exactly at any finite order, they
can be used as a lattice-independent criterion for a choice of the optimal subtraction point.

Since the total action has not been modified, the general renormalization of the theory is not affected by
the existence of the mass parameter $m$ which must be regarded as a RG invariant phenomenological
scale,  fixing the units of the scale-less theory. In that role, the mass parameter $m$
replaces the scale $\Lambda_{QCD}$ of the standard PT. Actually, the screened expansion and the standard
PT share the same general renormalization procedure and the same set of free parameters, since they are different
approximations for the same theory. That makes a great difference with other phenomenological models, like
the Curci-Ferrari model where a bare mass is added by hand and requires a mass renormalization constant.

The SMOM scheme emerges as the natural renormalization scheme of the screened expansion and the multiplicative
renormalization of the theory leads to a stationary condition which gives $\mu\approx m$ as the optimal
subtraction point, where the optimal strong coupling is predicted to be $\alpha_s(m)\approx 2.23$, corresponding
to $\alpha_s(m)\approx 1.12$ in the standard MOM scheme. 
It is remarkable that, once optimized from first principles, the screened expansion gives an excellent description of
the propagators and of the strong coupling if the explicit analytical results are compared with the lattice data in
the MOM scheme. Moreover, the lattice-independent measure of the coupling seems to be compatible with the value which
can be extrapolated from the lattice data in the continuum limit\cite{duarte}.

The SMOM scheme provides a quite appealing non-perturbative strong coupling which is a monotonic decreasing function of
the scale $\mu/m$ and saturates in the IR at the finite stable fixed point $\alpha_s(0)=7.03\approx 2.2\,\pi$,
in good agreement with previous estimates by functional RG\cite{gies} and gauge-gravity duality\cite{brodsky10,brodsky16}.
An explicit expression for the beta function is provided in the IR with the asymptotic behavior $\beta(g)=2(g-g_0)$ for
$g<g_0$. In order to make contact with the phenomenology, it would be interesting to extend the same scheme to the full QCD, 
including a quark loop in the gluon polarization and comparing the calculated coupling with experimental measures and
other approaches\cite{roberts,brodsky16R}.

An other important test would arise from a detailed study of the gauge dependence of the coupling in the SMOM scheme.
In order to be relevant for the phenomenology, the calculated strong coupling should be gauge invariant, but that would be
hard to tell by a one-loop calculation. The Landau gauge is quite special since $Z_1^c=1$, exactly, because of the 
non-renormalization theorem. In a generic covariant gauge, the SMOM scheme would require an explicit calculation of the
vertex for determining the finite part of $Z_1^c$, which would also depend on the gauge parameter $\xi$.
While the total gauge dependence might not cancel exactly in the coupling, the residual dependence might not be genuine as it
could arise from the further one-loop approximation for the vertex. In that respect, 
the Landau gauge provides our best estimate for the strong coupling.

\acknowledgments

The author is in debt to Orlando Oliveira for sharing the lattice data of Ref.\cite{duarte}.
This research was supported in part by "Piano per
la Ricerca di Ateneo 2017/2020 - Linea di intervento 2" of the University of Catania.

\appendix
\section{Explicit expressions}

\subsection{One-loop self-energies}

The one-loop self energies $F(x)$, $G(x)$ were first derived in Refs.\cite{ptqcd,ptqcd2} up to
third order in the $\delta$-expansion. The explicit expressions are
\begin{align}
F(x)&=\frac{5}{8x}+\frac{1}{72}\left[L_a+L_b+L_c+R\right]\nn\\
G(x)&=\frac{1}{12}\left[L_{gh}+R_{gh}\right]
\label{FGx}
\end{align}
where the logarithmic functions $L_x$ are
\begin{align}
L_a(x)&=\frac{3x^3-34x^2-28x-24}{x}\>\times\nn\\
&\times\sqrt{\frac{4+x}{x}}
\ln\left(\frac{\sqrt{4+x}-\sqrt{x}}{\sqrt{4+x}+\sqrt{x}}\right)\nn\\
L_b(x)&=\frac{2(1+x)^2}{x^3}(3x^3-20x^2+11x-2)\ln(1+x)\nn\\
L_c(x)&=(2-3x^2)\ln(x)\nn\\
L_{gh}(x)&=\frac{(1+x)^2(2x-1)}{x^2}\ln(1+x)-2x\ln(x)
\label{logsA}
\end{align}
and the rational parts are
\begin{align}
R(x)&=\frac{4}{x^2}-\frac{64}{x}+34\nn\\
R_{gh}(x)&=\frac{1}{x}+2.
\label{rational}
\end{align}

The derivative of $F(x)$ is\cite{xigauge}
\BE
F^\prime (x)=-\frac{5}{8x^2}+\frac{1}{72}\left[L^\prime_a+L^\prime_b+L^\prime_c+R^\prime\right]
\label{der1}
\EE
where the logarithmic functions $L^\prime_x$, for $x=a,b,c$, are
\begin{align}
L^\prime_a(x)&=\frac{6x^4-16x^3-68x^2+80x+144}{x^2(x+4)}\>\times\nn\\
&\times\sqrt{\frac{4+x}{x}}
\ln\left(\frac{\sqrt{4+x}-\sqrt{x}}{\sqrt{4+x}+\sqrt{x}}\right)\nn\\
L^\prime_b(x)&=\frac{4(1+x)}{x^4}(3x^4-10x^3+10x^2-10x+3)\ln(1+x)\nn\\
L^\prime_c(x)&=-6x \ln x
\end{align}
and $R^\prime(x)$ is the sum of all the rational terms coming out from the derivatives
\BE
R^\prime(x)=\frac{12}{x}+\frac{106}{x^2}-\frac{12}{x^3}.
\EE
Observe that the functions $L^\prime_x$, $R^\prime$ are not the derivatives of the corresponding functions $L_x$, $R$.

\subsection{IR leading behavior}

Deep in the IR, the following leading behavior is obtained in the limit $x\to 0$
\begin{align}
L_a(x)&=\frac{48}{x}+60+\frac{7\cdot 13}{15}\,x+{\cal O}(x^2)\nn\\
L_b(x)&=-\frac{4}{x^2}+\frac{16}{x}+44-\frac{157}{3}+\frac{103}{3}\,x+{\cal O}(x^2)
\end{align}
yielding for the function $F(x)$
\begin{align}
F(x)&=\frac{5}{8x}+\frac{257}{3\cdot 72}+ \frac{101}{5\cdot 36}\,x+\frac{1}{36}\,\ln x+{\cal O}(x^2)\nn\\
&=\frac{c}{x}+c_0+ c_1\,x+c_L\,\ln x+{\cal O}(x^2)
\end{align}
which has been checked to be a very good approximation for $x<0.01$ ($p< 0.1\,m$).
At the same order of approximation, the dressing function $J(t^\star,x)$ can be written as
\begin{align}
J(t^\star,x)&=\frac{(1+x^{-1})}{F(x)+F_0(t^\star)}\nn\\
&\approx\frac{1+x}{c+(c_0+F_0)\,x+ c_1\,x^2+c_L\,x\,\ln x}\nn\\
&\approx \frac{1}{c}\left[1+(a_J-b_J\,\ln x)x\,\right]
\end{align}
which holds for $x<0.015$, with the constants:
\begin{align}
c&=\frac{5}{8}=0.625,\quad c_0=\frac{257}{3\cdot 72}\approx 1.190\nn\\
a_J&=1-\frac{c_0+F_0}{c}\approx 0.498\quad (F_0=-0.876)\nn\\
b_J&=\frac{c_L}{c}=\frac{8}{5\cdot 36}\approx 0.0444.
\end{align}

The leading behavior of the function $L_{gh}(x)$ is
\BE
L_{gh}(x)=-\frac{1}{x}+\frac{1}{2}+\frac{8x}{3}-2x\ln x+{\cal O}(x^2)
\EE
yielding the approximate expression
\BE
G(x)+G_0\approx k_0+x\,(k_1-k_L\,\ln x)
\EE
which has been checked to be a very good approximation for $x<0.2$,
with the constants
\BE
k_0=\frac{5}{24}+G_0,\quad k_1=\frac{2}{9},\quad k_L=\frac{1}{6}.
\EE
At the same order of approximation, the dressing function $\chi(t^\star,x)$ can be written as
\begin{align}
\chi(t^\star,x)&=\frac{1}{G(x)+G_0(t^\star)}
\approx \frac{1}{k_0}\left[1-(a_\chi-b_\chi\,\ln x)x\,\right]
\end{align}
which holds for $x<0.01$, with the constants:
\begin{align}
k_0&=\frac{5}{24}+G_0(t^\star)\approx 0.3536 \quad (G_0=0.14524)\nn\\
a_\chi&=\frac{k_1}{k_0}\approx 0.6285\nn\\
b_\chi&=\frac{k_L}{k_0}\approx 0.47138.
\end{align}

By insertion in Eq.(\ref{SMOMrun}), deep in the IR, the leading behavior of the SMOM running coupling is

\begin{align}
g(x)&=g_0\,\left[1+(a-b\,\ln x)x\,\right]+{\cal O}(x^2)
\label{glead}
\end{align}
where
\begin{align}
a&=\frac{1}{2}a_J-a_\chi\approx -0.3795\nn\\
b&=\frac{1}{2}b_J-b_\chi\approx -0.449
\end{align}
and $g_0$ is the IR stable fixed point
\BE
g_0=\frac{4\pi}{\sqrt{3\,c\,N}}\>\frac{1}{k_0\,\alpha(t^\star)}\approx 9.4017 
\label{g0}
\EE
for $\alpha(t^\star)=1.5939$ and $N=3$.

The leading behavior of the beta function in the IR follows by its definition
\BE
\beta(g)=\mu\,\frac{{\rm d} g}{{\rm d} \mu}=2x\,\frac{{\rm d} g}{{\rm d} x}
\EE
which by Eq.(\ref{glead}) gives the coupled equations
\begin{align}
\beta(g)&=2(g-g_0)-2b\,g_0\,x(g)\nn\\
x(g)&=f^{-1}\left(\frac{g-g_0}{g_0}\right)
\label{betaimpl}
\end{align}
where $f(x)=x\,(a-b\,\ln x)$.
At the same order of approximation, the function $f(x)$ can be inverted by iteration.
Denoting by $\epsilon$ the small positive variable
\BE
\epsilon=\frac{g-g_0}{b\,g_0}\to 0\quad {\rm for}\quad x\to 0
\EE
and observing that $f[x(g)]=b\,\epsilon$, the function $f(x)$ can be written as
\BE
x(g)=\frac{f[x(g)]}{a-b\,\ln x(g)}=\frac{\epsilon}{\eta-\ln x(g)}
\EE
where $\eta=a/b$. Then, iterating,
\begin{align}
x(g)&=\>\epsilon\left[\eta-\ln\epsilon+\ln(-\ln x) +\ln\left(1-\frac{\eta}{\ln x}\right)\right]^{-1}\nn\\
=&\>\epsilon\left[\eta+\ln\frac{1}{\epsilon}+\ln\left(\ln \frac{1}{\epsilon}\right) + {\cal O}\left(\frac{1}{\ln (1/\epsilon)}\right)
\right]^{-1}
\end{align}
and inserting in Eq.(\ref{betaimpl}) we obtain an explicit expression for the leading beahvior of the beta in the IR
\begin{align}
\beta_{IR}&
\approx 2bg_0\,\epsilon\left\{1-\left[\eta+\ln\frac{1}{\epsilon}+\ln\left(\ln\frac{1}{\epsilon}\right)\right]^{-1}\right\}.
\label{betaexpl}
\end{align}

\end{document}